\documentclass{article}
\usepackage{epsfig}
\newcommand{\bfr}{\begin{flushright}}
\newcommand{\efr}{\end{flushright}}
 
\begin{document}
% \eqsec  % uncomment this line to get equations numbered by (sec.num)
\title{Spinning a charged dilaton black hole
%\thanks{Presented at ...}%
% you can use '\\' to break lines
}
\author{Kiyoshi Shiraishi\\
%\address{
Akita Junior College, Shimokitade-Sakura, Akita-shi, \\Akita 010,
Japan%}
}
\date{Physics Letters A 166 (1992) 298--302
}
\maketitle
\begin{abstract}
A charged dilaton black hole which possesses infinitesimal
angular momentum is studied. We find that the gyromagnetic ratio of the
dilaton black hole depends not only on the parameter which appears in
the interaction between the dilaton and the electric field but also
nonlinearly on the ratio of the charge to the mass of the black hole.
The moment of inertia for the charged dilaton hole in the limit of
infinitesimal angular momentum is also calculated.
\end{abstract}
%\PACS{}

\bigskip

Exact solutions for
charged black holes in effective field theory of string theory have
been constructed and reexamined recently \cite{1,2,3}. It is pointed out
that the dilaton plays an important role in these solutions. The
validity of a thermodynamical description for these black holes is
argued by many authors \cite{4,5}. In the case of charged dilaton holes
in low-energy string theory, the laws of black hole thermodynamics come
into question. An interpretation is suggested that the charged dilaton
hole in string theory has the properties of an elementary particle,
rather than a thermal state. 

Modification of the dilaton coupling to an
electromagnetic field has been examined in general cases and the
analysis reveals that both the standard Reissner-Nordstr\"om black
hole and the charged black hole in string theory are very special cases.
(Black hole solutions in models including axion coupling and/or
gravitational higher-derivative terms are obtained in refs.
\cite{6,7,8} and references therein. In the present Letter we do not
treat the awkward entries.) 

As a next step to understand the property
of dilaton holes, we wish to look into another hair of the black hole:
angular momentum. Unfortunately, exact solutions to the Einstein
equation coupled to general matter fields are unknown in most cases;
only for limited cases, an exact solution is obtained. An exact
solution for a rotating (charged) black hole with a special dilaton
coupling is derived using the inverse scattering method \cite{9} (for
neutral rotating solitons see ref. \cite{10}). Since these solutions are
constructed by use of dimensional reduction, a similar method is not
available for the case with general dilaton coupling. 

In this Letter,
for these reasons, we study the charged dilaton black hole with
infinitesimal angular momentum and try to extract the characteristics
of the rotating hole. 

We start with the action 
\begin{equation}
S=\int d^4x [R-2(\nabla\phi)^2-\exp(2a\phi)F^2]\,,
\label{eq1}
\end{equation}
where $a$ is the parameter which
determines the strength of the coupling between the Maxwell field $F$ and
the dilaton field $\phi$. For $a=0$, the action reduces to the one which
governs the usual Einstein-Maxwell system with a free scalar field. A
charged spherical black hole is described by the Reissner-Nordstr\"om
solution in the system. For $a=1$, the action coincides with the one
which is derived from low-energy string theory. 

The solution for a
charged spherical black hole in the system governed by the general
action (\ref{eq1}) can be written in the form \cite{1,2}
\begin{equation}
ds^2=-f\,dt^2+f^{-1}dr^2+R^2(r)(d\theta^2+\sin^2\theta d\varphi^2)\,,
\label{eq2}
\end{equation}
where
\begin{equation}
f=(1-r_{+}/r)(1-r_{-}/r)^{(1-a^2)/(1+a^2)}\,,
\label{eq3}
\end{equation}
and
\begin{equation}
R^2=r^2(1-r_{-}/r)^{2a^2/(1+a^2)},.
\label{eq4}
\end{equation}
In this solution $r_+$ and $r_-$ are integration constants and they are
connected to the physical quantities of the black hole through the
relations 
\begin{eqnarray}
2M&=&r_++\frac{1-a^2}{1+a^2}r_-\,,\\
Q^2&=&\frac{r_-r_+}{1+a^2}\,,
\end{eqnarray}
where $M$ and $Q$ are the mass and the electric charge of the black
hole, respectively. The horizon is located at $r=r_+$, and the maximal
charge is attained when $r_+=r_{-}$, i.e., $Q_{max}=(1+a^2)^{1/2}M$.

The dilation and the electric field behave as
\begin{equation}
\exp(2a\phi)=(1-r_{-}/r)^{2a^2/(1+a^2)}\,,
\label{eq7}
\end{equation}
and
\begin{equation}
\exp(-2a\phi)F=\frac{Q}{R^2} dt\wedge dr 
\label{eq8}
\end{equation}
(or, equivalently, $A_t=Q/r$).

If an infinitesimal angular momentum $j$ (parallel to the $z$-axis) is added, we need only
to know a few components of the gauge field and the metric. They are $A_\phi$
and $g_{t\phi}$, which are of the order of $j^1$. The corrections
to the other components of the field and the metric are $O(j^2)$, and
ignored in the present analysis. Note that the equation of motion for
the dilaton gets no correction at the first order and thus the behavior
of the dilaton is unchanged from (\ref{eq7}). Therefore we find that $j^2/M^2$
cannot be as much as $Q^2$. 

The asymptotic form of the field components is
given by 
\begin{eqnarray}
A_\phi&=&\mu\frac{\sin^2\theta}{r}+O(1/r^2)\,,\label{eq9}\\
g_{t\phi}&=&2j\frac{\sin^2\theta}{r}+O(1/r^2)\,,
\label{eq10}
\end{eqnarray}
where the constant $\mu$ is interpreted as the magnetic moment.

We are looking for the solution which behaves as (\ref{eq9}) and (\ref{eq10}) at
infinity. Solving the linearized equation of motion derived from the
action (\ref{eq1}), we obtain 
\begin{equation}
A_\phi=Q\alpha\frac{\sin^2\theta}{r}\,,
\label{eq11}
\end{equation}
and
\begin{eqnarray}
& &g_{t\phi}=\alpha\sin^2\theta\left[\frac{r^2}{r_-^2}\frac{1+a^2}{1-a^2}\left(
\frac{1+a^2}{1-3a^2}
[(1-r_{-}/r)^{(3a^2-1)/(1+a^2)}-1]\right.\right.\nonumber \\
& &\left.\left.-\frac{r_-}{r}\right)-(1-r_+/r)\right]
(1-r_{-}/r)^{(1-a^2)/(1+a^2)}\,, \quad\mbox{for } a^2\ne 1, a^2\ne 1/3\,, \label{eq12}\\
&
&g_{t\phi}=\alpha\sin^2\theta\left(\frac{2r^2}{r_-^2}[r_-/r+(1-r_{-}/r)
\ln(1-r_{-}/r)]-(1-r_+/r)\right)\,,
\quad\mbox{for }  a^2=1\,, \label{eq13}\\ 
& &g_{t\phi}=\alpha\sin^2\theta\left(-\frac{2r^2}{r_-^2}
[r_-/r+\ln(1-r_{-}/r)]-(1-r_{+}/r)\right) (1-r_{-}/r)^{1/2}\,, \quad\mbox{for }  a^2=1/3\,,
\label{eq14}
\end{eqnarray}
where $\alpha$ is a constant.

Using the asymptotic behaviors, we find that the gyromagnetic ratio is given by
\begin{equation}
\frac{\mu}{j}=\frac{Q}{M}\frac{(1+a^2)r_++(1-a^2)r_-}{(1+a^2)r_++(1-a^2/3)r_{-}}
\equiv g\frac{Q}{2M}\,.
\label{eq15}
\end{equation}
For $a=0$, (\ref{eq15}) reduces to the well-known result for a Kerr-Newman
black hole \cite{11}: $g=2$ (like an electron!). For finite $a$, $g$ is
expressed as a function of $Q/M$. $g$ is smaller than $2$ for finite $Q/M$ in
general, whereas the relation 
\begin{equation}
\lim_{Q/M\rightarrow 0} g(Q/M)=2
\label{eq16}
\end{equation}
is satisfied for any value of $a$. The dependence on $Q/M$ for some cases is shown 
in fig.~1. 

Before discussing the physical implication, we will investigate another physical
quantity. 

%**************************************************************
\begin{figure}[ht]
\begin{center}
\includegraphics[width=6cm]{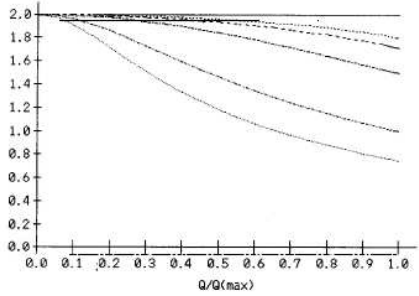}
\caption{Fig. 1. The $g$ factor of the charged dilaton black hole with small
angular momentum versus the charge-to-mass ratio of the black hole.
(-----) $a^2=0$; (- - - -) $a^2=1/3$; (-- -- --) $a^2=1/2$; (-- $\cdot$ --) $a^2=1$; (--
$\cdot$  $\cdot$ --)
$a^2=3$; ($\cdots\cdots$)
$a^2=5$.}
\label{f1}
\end{center}
\end{figure}
%**************************************************************

The angular velocity at the horizon $r=r_+$ is given in the leading order by
\begin{equation}
\Omega_+\equiv\frac{g_{t\phi}(r=r_+, \theta=\pi/2)}{R^2(r_+)}\,. 
\label{eq17}
\end{equation}
The explicit expression for $\Omega_+$ can be obtained by using (\ref{eq12})--(\ref{eq14}).

The moment of inertia for the charged dilaton hole with infinitesimal angular momentum, 
$I=j/\Omega_+$,is exhibited
in fig.~2. We see, for any value of $a$, that
\begin{equation}
\lim_{Q/M\rightarrow 0}\frac{I}{MR^2(r_+)}=1\,.
\label{eq18}
\end{equation}
The critical value for $a^2$ is $1$. For $a^2<1$, $I/MR^2(r_+)$ diverges at $Q=Q_{max}$;
for $a^2>1$, $I/MR^2(r_+)$ approaches zero when $Q\rightarrow Q_{max}$. For $a^2=1$,
$I/MR^2(r_+)$ diverges logarithmically near $Q=Q_{max}$.

%**************************************************************
\begin{figure}[ht]
\begin{center}
\includegraphics[width=6cm]{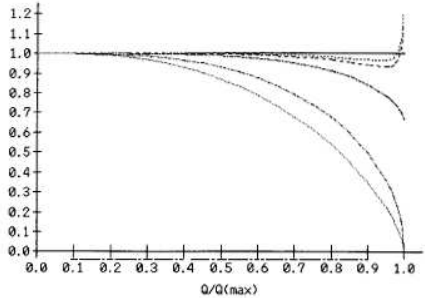}
\caption{Fig. 2. The moment of inertia, $I$, for the charged dilaton black hole
with small angular momentumversus the charge-to-mass ratio ofthe black
hole. The moment of inertia $I$ is normalized by $MR^2(r_+)$. Designations as
in fig~1.}
\label{f2}
\end{center}
\end{figure}
%**************************************************************

Now we are working with the assumption of infinitesimal rotation; we must notice that the
present result is not very trustworthy in the extremal limit, unfortunately, since even
small rotation may change the extremity condition.

Except for the region where $Q\sim Q_{max}$, $I/MR^2(r_+)$ is smaller than 1. This result,
when combined with the previous result $g<2$, indicates that the ratio of the magnetic
moment to the angular velocity decreases when $a^2 \ne 0$ and $Q^2$ is finite. Therefore
one can say that the ``charge distribution'' must be changed by the configuration of the
dilaton. (Here we wish to say that if the charge distribution of an object remains
unchanged, the reduction of the moment of inertia leads to an increase of the gyromagnetic
ratio of the object according to classical electromagnetism.) 

Actually an interpretation
is possible in the case with small $Q$ (i.e., $r_-/r_+ \ll 1$). The charge
``observed'' at the horizon is deduced from the Gauss law and (\ref{eq7}), (\ref{eq8}):
\begin{equation}
Q_{obs}/Q=(1-r_{-}/r_+)^{2a^2/(1+a^2)}\sim 1-\frac{2a^2}{1+a^2}\frac{r_-}{r_+}\,.
\label{eq19}
\end{equation}
If the ``observed'' charge at the horizon induces the magnetic moment
of the hole, the $g$ factor must take the above correction. Compared with
(\ref{eq15}), unfortunately, the explanation is only a qualitative one (in
spite of all the elaboration of the other factors). 

In summary, we have
managed to give a charged dilaton hole a small angular momentum. We
have found that the $g$ factor of the black hole depends not only on a
but also on $Q/M$. The moment of inertia for the black hole has also
been obtained in the limit of infinitesimal angular momentum.

We have found that both the $g$ factor and the (normalized) moment of
inertia of the black hole decrease as the parameter a increases. Both
quantitites decrease also as the charge of the hole increases in most
of the parameter region except for $Q\sim Q_{max}$. 

Quantum processes
associated with the curved space near the dilaton black hole may well
be worth studying. Incidentally, the angular velocity ofthe dilaton
black hole obtained here is relevant to the phenomenon of super
radiance \cite{12}. The future study of quantum processes will clarify
various aspects of the thermodynamics of black holes. 

\bigskip

\noindent
{\bf Note added.} After
the submission of the present paper, the author has been informed
of refs.~\cite{13} and \cite{14}. Ref.~\cite{13} treats a similar study
while ref.~\cite{14} claims that the effective action (\ref{eq1}) with $a=1$
does not exactly correspond to that of string theory because of the
excitation ofan antisymmetric field strength by the non-zero electric
and magnetic field via $dH=F\wedge F$. 

I should like to thank the referee for
this important information.

%%%%%%%%%%%%%%%%%%%%%%%%%%%%%%%%%%%%%%%%%%%%%% 

%%%%%%%%%%%%%%%%%%%%%%%%%%%%%%%%%%%%%%%%%%%%%
\end{document}